\documentclass[12pts,twocolumn,epsf,aps,prb]{revtex4}
\usepackage{color,graphicx}
\begin{document}

\title{Memory effects and magnetic relaxation in single-crystalline La$_{0.9}$Sr$_{0.1}$CoO$_3$}

\author{N. Khan$^1$, P. Mandal$^1$ and D. Prabhakaran$^2$}

\affiliation{$^1$Saha Institute of Nuclear Physics, 1/AF Bidhannagar, Kolkata 700 064, India}
\affiliation{$^2$Department of Physics, Clarendon Laboratory, University of Oxford, Oxford OX1 3PU, United Kingdom}

\date{\today}
\begin{abstract}

We report a detailed investigation of magnetic relaxation and memory effects in La$_{0.9}$Sr$_{0.1}$CoO$_3$ single crystal from dc magnetization measurements. The analysis of magnetic relaxation at different temperatures and magnetic fields below the freezing temperature ($T_{f}$) manifests the characteristics of spin glass phase. Below $T_{f}$, striking memory effects have been clearly observed in different experimental protocols. The memory effect in the zero-field-cooled magnetization further establishes that the glassy magnetic state arises from the co-operative spin-spin interaction but not due to the independent relaxation of metastable phase clusters. The asymmetric response with respect to negative and positive temperature change favors the hierarchical model of memory effects rather than the droplet model discussed in other works for different insulating and metallic Heisenberg spin glasses.
\\

{PACS number(s):75.40.Cx,75.50.Lk,75.10.Nr,75.47.Lx}
\end{abstract}
\maketitle
\newpage
\section{INTRODUCTION}
Spontaneous magneto-electronic phase separation (MEPS) has been extensively studied in different complex oxides. MEPS  is believed to be the key ingredient to understand the underlying mechanism responsible for the intriguing physical properties in manganites and high temperature cuprate superconductors.\cite{dagotto,edagotto} Unlike manganite, the hole-doped La$_{1-x}$Sr$_{x}$CoO$_3$ (LSCO) cobaltite is a model system to study the MEPS phenomenon because of the absence of charge ordering, insulating ferromagnetic ordering as well the long-range antiferromagnetic ordering, which make it easier to probe and understand the phase separation phenomenon.\cite{che,he} Substitution of divalent Sr$^{2+}$ ions at trivalent La$^{3+}$ sites in LaCoO$_3$ causes spontaneous nanoscopic phase separation where nano-sized (i.e., 1-3 nm) hole-rich ferromagnetic (FM) metallic clusters  are embedded in a hole-poor insulating non-FM matrix.\cite{che,he,dphelan} The interaction between Co$^{4+}$ and Co$^{3+}$ is  ferromagnetic double exchange whereas that between Co ions with same valance states (Co$^{3+}$$-$Co$^{3+}$ and Co$^{4+}$$-$Co$^{4+}$) is  antiferromagnetic superexchange.\cite{senaris,bhide,ganguly} For low doping, these two competitive interactions are random and frustrated which lead to the glassy magnetic behavior for the doping range 0.0$<$$x$$<$0.18.\cite{senaris,itoh,mitoh,wu} With increasing doping level, the number and size of the hole-rich FM clusters increase rapidly and the percolation of these FM clusters at a critical doping value, $x_{p}$$\simeq$0.18, yields a crossover from a short-range to long-range FM ordering.\cite{che,he,dphelan} Accumulated experimental data from various high-resolution probes reveal that the phase separation in LSCO crystals is confined to a well defined doping range, 0.04$<$$x$$<$0.22, which covers both the FM and glassy magnetic states.\cite{che,he} It is fascinating to investigate whether the phase separated states in cobaltites exhibiting glassy magnetic behavior resemble that of a canonical/atomic spin glass or a superparamagnetic system or an assembly of strongly interacting magnetic clusters. Earlier studies on polycrystalline compounds (0$<$$x$$<$0.15) showed the presence of multiple glassy magnetic phases at low temperatures.\cite{senaris,itoh,wu} It was observed that those compounds exhibit superparamagnetic behavior originating from isolated/non-interacting FM clusters below the irreversibility temperature ($T_{irr}$) at which bifurcation of the field-cooled (FC) and zero-field-cooled (ZFC) magnetizations occurs. As the temperature is lowered below $T_{irr}$, frustrated intercluster interaction develops and results in a blocking of these superparamagnetic clusters at a characteristic spin blocking temperature ($T_{g}$). Both the $T_{irr}$ and $T_{g}$ were found to exhibit doping dependence, where the former decreases and the later increases with increasing hole doping.\cite{senaris,wu} Contrary to this, in single-crystalline compounds with $x$=0.10 and 0.15 and a highly homogeneous $x$=0.05 polycrystalline compound, the bifurcation of the ZFC and FC magnetization occurs close to spin freezing temperature $T_{f}$ and no such $T_{irr}$ was found and they exhibit characteristics very similar to spin glass phase.\cite{nkhan,manna,kmanna,nam} Even at percolation threshold ($x$=0.18), the polycrystalline compound exhibits glassy magnetic behavior and ageing effect which are believed to be originating from both spin-glass-like phase and interacting FM clusters\cite{ktang}, though the single-crystalline LSCO exhibits long-range ferromagnetic behavior with the Curie temperature $T_{C}$=150 K.\cite{tang,Ewings} Therefore, the origin of several glassy magnetic behavior in LSCO compounds for doping 0$<$$x$$<$0.18 is still not settled unambiguously. Compositional inhomogeneity in the polycrystalline samples could be reason for such ambiguous glassy magnetic behavior\cite{samal} and  responsible for the occurrence of MEPS up to doping level as high as $x$=0.5.\cite{kuhns,rhoch,hoch} The inhomogeneous distribution of Sr$^{2+}$ ions results in a distribution of ferromagnetic cluster sizes which dominates the characteristic features of magnetism and may mask the intrinsic magnetic properties of the stoichiometric phase. Therefore, to resolve those long-standing ambiguities and to know the intrinsic magnetic ground state one should perform a through investigation of different glassy magnetic behavior on high quality single-crystalline samples. Also glassy behavior such as fascinating memory effects and ageing properties are believed to be of great practical use as have been recently investigated in a large number of experiments on magnetic nanoparticles.\cite{Sun,Tsoi,Zheng,Kundu,Sasaki,De,Pradheesh} As the present system for low doping (i.e., $x$$<$0.15) phase separates into nanoscopic FM droplets in the background of non-FM matrix, the magnetic freezing process in this system is conceptually similar to that of magnetic nanoparticles.\cite{Aarbogh} Therefore, such properties should also be investigated in greater details to tailor this material for possible technological applications. Recently, we have studied the critical behavior of the spin glass freezing transition for the La$_{0.9}$Sr$_{0.1}$CoO$_3$ single-crystalline compound which is very close to the middle of the so called spin glass regime and  showed the spin glass like phase of the compound.\cite{nkhan} In the present report, we have investigated its magnetic relaxation at different temperatures and magnetic fields which is an another experimental method to distinguish spin glass from that of cluster glasses and supermagnetic systems.\cite{Rivadulla,Ulrich} Indeed, our analysis of magnetic relaxations below the freezing temperature following the model proposed by Ulrich \emph{et al.}\cite{Ulrich} clearly manifests a  crucial difference between the phase separated state of glassy manganite and cobaltite systems. Besides, we have performed a detailed study on the fascinating memory effects using different experimental protocols that reveals  important differences between different glassy phases.\cite{malay} The present compound is also found to exhibit quite strong memory effects and rejuvenation which have been discussed in the light of the two well-known phenomenological scenarios viz., the droplet \cite{Fisher,Yoshino} and hierarchical models \cite{Evincent,Dotsenko,Lefloch,Vincent} and compared with those of insulating and metallic Heisenberg spin glasses\cite{hMamiya,rMathieu,Mathieu}.
\\
\section{EXPERIMENTAL DETAILS}

The single-crystalline La$_{0.9}$Sr$_{0.1}$CoO$_3$ was prepared by the traveling solvent float zone method using a four mirror image furnace (Crystal System Inc.) The single phase and high quality crystalline nature were confirmed by different experimental techniques such as x-ray powder diffraction, Laue diffraction, scanning electron microscopy (SEM), and electron probe microanalysis. The details of synthesis and different characterizations have been reported elsewhere.\cite{prabhak,mandal} All the measurements in the present study were performed using highly sensitive SQUID VSM system (Quantum Design). The memory effects and rejuvenation measurements were performed using different experimental protocols \cite{Sun} and the magnetic relaxation measurements were carried out by conventional procedure.\\

\section{Results and Discussion}

\begin{figure}
\includegraphics[width=0.5\textwidth]{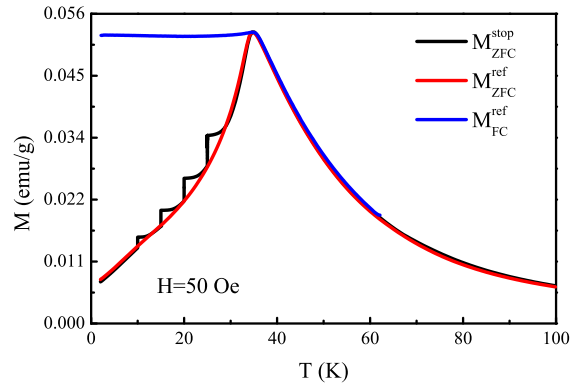}
\caption{(Color online) (a) Temperature dependence of the field-cooled ($M_{FC}^{ref}$) and zero-field-cooled ($M_{ZFC}^{ref}$) dc magnetization at 50
Oe for the La$_{0.9}$Sr$_{0.1}$CoO$_3$ single crystal. The step-like zero-filed-cooled magnetization ($M_{ZFC}^{stop}$) is due to the intermediate halts of duration 1 h each at temperatures 10, 15, 20, and 25 K.}
\label{Figure.1}
\end{figure}
Figure 1 depicts the temperature dependence of field-cooled magnetization ($M_{FC}^{ref}$) and zero-field-cooled magnetization ($M_{ZFC}^{ref}$) measured by conventional procedure as well as the temperature dependence of zero-field-cooled magnetization ($M_{ZFC}^{stop}$) measured with intermediate halts at different stopping temperatures of duration 1 h each. All the magnetization measurements were recorded in the warming cycle for an applied dc field of 50 Oe. The  step-like $M_{ZFC}^{stop}$ curve shows a clear time evolution of the magnetization at various stopping temperatures and suggests glassy dynamics below the freezing temperature $T_{f}$=34.8 K. The ageing effect and the time dependence of magnetization below $T_{f}$ have been investigated by measuring magnetic relaxations using different experimental protocols. Figure 2 shows the relaxation of the zero-field-cooled magnetization at 5 K for an applied field of 10 Oe after different wait times $t_{w}$=100, 1000, and 5000 s in zero field. For each relaxation curve, the sample was initially cooled in zero field from a reference temperature 100 K in the paramagnetic state to a measuring temperature $T_{m}$=5 K, which is well below the freezing temperature and kept at $T_{m}$ for the respective wait time. After the lapse of $t_{w}$, 10 Oe dc field was applied and the time evolution of the magnetization was recorded as shown in Fig. 2. The time dependence of the ZFC magnetization can be fitted well with the stretched exponential function of the following form
\begin{figure}
\includegraphics[width=0.5\textwidth]{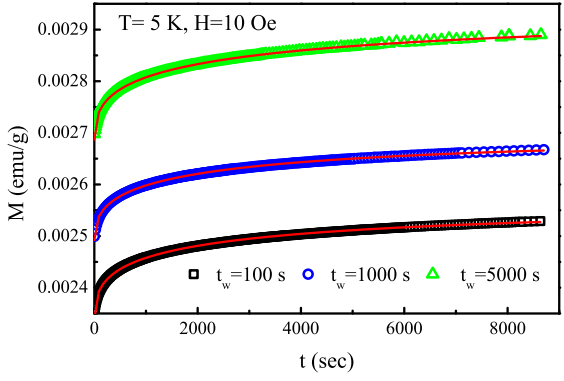}
\caption{(Color online) Time dependence of magnetization at 5 K measured in the ZFC state with an applied field of 10 Oe for different waiting time ($t_{w}$=100, 1000, and 5000 s). The solid lines are the best fit to the stretched exponential function following Eq. (1)}.
\label{Figure.2}
\end{figure}
\begin{equation}
M(t) = M_{0}-M_{r}exp\{-(t/\tau_{r})^{\beta}\},
\end{equation}
where $M_{0}$  is an intrinsic FM component. $M_{r}$ and $\tau_{r}$ are the glassy component and characteristic relaxation time, respectively, both of which depend on $T$ and $t_{w}$. $\beta$ is called the stretching parameter which lies between 0 and 1 and is a function of temperature only. The estimated parameters of the magnetic relaxation for the present compound are listed in Table I. The deduced value of $\beta$$\approx$0.42 is found to lie within the range reported for different glassy systems and is close to that of the canonical spin glass [Cu:Mn(6\%)].\cite{chu} Also, $\beta$$<$1 signifies that the system evolves through a number of intermediate states i.e., activation takes place against multiple anisotropic barriers. The  value of the time constant $\tau_{r}$ increases with the increase of $t_{w}$, manifesting the stiffening of the spin relaxation or the ageing effect.
\begin{table*} {
\caption{Fitted parameters of the magnetic relaxations at 5 K for different wait times described in the Fig. 2 using Eq. (1).}
\label{I}
\begin{tabular*}{1.0\textwidth}{@{\extracolsep{\fill}}c c c c c }
\hline 
 $t_{w}$(s) & $M_{0}$(emu/g) & $M_{r}$(emu/g) & $\tau_{r}$(s) & $\beta$\\
\hline
100  & 0.0026(1) & 2.23(2)$\times$10$^{-4}$ & 1947(46) & 0.414(4) \\[6pt]
1000  & 0.0027(1) & 2.13(1)$\times$10$^{-4}$ & 2286(48) & 0.418(2) \\[6pt]
5000  &0.0029(3) & 2.52(5)$\times$10$^{-4}$ & 2802(161) & 0.415(7) \\[6pt]
\hline
\hline
\end{tabular*}}
\end{table*}\\
\\
The magnetic relaxation rate has been analyzed using the theoretical model proposed by Ulrich \emph{et al.}\cite{Ulrich} which has been used earlier to study the relaxation behavior of manganite.\cite{Rivadulla} According to this model, for an assembly of interacting magnetic particles, the relaxation rate, $W(t)$=$-$(d$/$d$t$) $ln$ $M(t)$, decays by the following power law after the lapse of a crossover time, $t_{0}$,
\begin{equation}
W(t)=At^{-n}
\end{equation}
\begin{figure}
\includegraphics[width=0.5\textwidth]{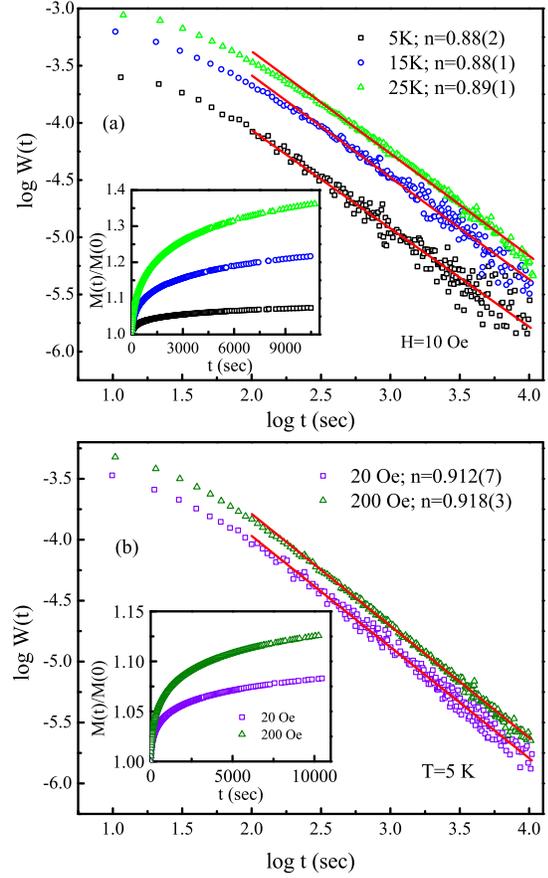}
\caption{(Color online) (a) The magnetic relaxation rate as a function of time on a log-log plot at 5, 15, and 25 K obtained from the corresponding ZFC magnetic relaxations measured in 10 Oe field as shown in the inset. The straight lines are due to the linear fit following Eq. (2) and the values of the exponent $n$ are obtained from the slopes of the fit. (b) Time dependence of the magnetic relaxation rate on log-log plot at 5 K for different applied magnetic fields ($H$=20 and 200 Oe) obtained from the corresponding ZFC magnetic relaxations as shown in the inset. The slopes of the linear fits to the relaxation rate give the values of the exponent $n$ for different fields.  For clarity, time dependence of relaxation rate at 100 Oe is not shown.}
\end{figure}
where $A$ is a constant and the exponent $n$ is a function of temperature and particle density. The value of $n$ gives a measure of the strength of dipolar interaction among the magnetic clusters.  Figure 3(a) depicts the relaxation rates $W$, obtained from the ZFC magnetic relaxation curves taken at 10 Oe dc field [inset of Fig. 3(a)], as a function of $t$ on log-log plot for different temperatures. The solid line is the least-square fit to Eq. (2) after a crossover time $t_{0}$$\simeq$10$^{2}$ s. The estimated values of $n$ are 0.88(2), 0.88(1), and 0.89(1) obtained from the magnetic relaxations at 5, 15, and 25 K, respectively. The field dependence of magnetic relaxation rate at a constant temperature is also investigated [Fig. 3(b)]. The time dependence of magnetic relaxation rate at 5 K is measured for different magnetic fields $H$=20, 100, and 200 Oe and the corresponding estimated values of $n$ are 0.912(7), 0.913(5), and 0.918(3), respectively. The important observation from the above analysis is that the estimated value of $n$ is found to be independent of temperature as well as magnetic field. On the contrary,  the value of $n$ in manganites increases continuously as the freezing temperature is approached from below, reflecting the increase of magnetic interaction between the clusters.\cite{Rivadulla} The value of the exponent $n$ increases with the increase of temperature as well as magnetic field for an assembly of interacting magnetic clusters with a distribution of particle size, whereas it remains constant for a true spin glass and a system of strongly interacting magnetic clusters of fixed size and density.\cite{Rivadulla} So, it appears that the cobaltite exhibiting glassy behavior resembles that of a true spin glass phase whereas the glassy behavior of the manganite mirrors the collective behavior that originates from the intercluster interactions only. Therefore, though both the cobaltite and manganite exhibit phase separation, the origins of glassiness are quite different for the two systems. Further, for the homogeneous $x$$=$0.15 polycrystalline compound which is below but close to the percolation threshold $x_{p}$, $n$ is also found to be independent of temperature  and its value matches very well with that estimated for the present single-crystalline sample.\cite{samal} This fact implies that the intrinsic magnetic ground state of the LSCO is canonical spin glass in the doping range 0.05$\leq$$x$$\leq$0.15.\\
\\
\begin{figure}
\includegraphics[width=0.5\textwidth]{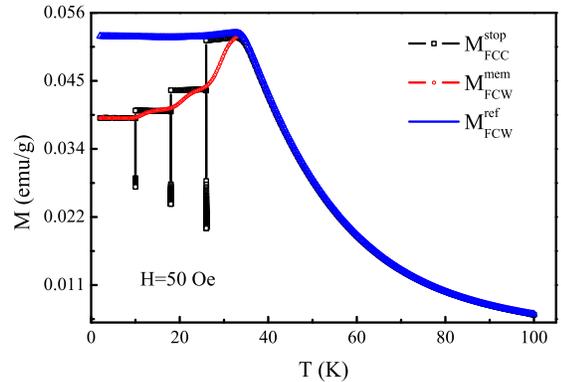}
\caption{(Color online) Memory effect in the temperature dependence of the field-cooled magnetization. $M_{FCW}^{ref}$ is the conventional or the reference field-cooled magnetization curve taken in 50 Oe field. The step-like curve $M_{FCC}^{stop}$ is obtained by measuring the magnetization during cooling with intermediate stops at 10, 18, and 26 K where the field is cut off for duration 2 h each after which field is again turned on and cooling is resumed. The magnetization curve $M_{FCW}^{mem}$ is called as the memory curve obtained during subsequent continuous reheating.}
\label{Figure.4}
\end{figure}
Figure 4 shows the memory effect in the thermal variation of the field-cooled (FC) magnetization obtained following the protocol of Sun \emph{et al.}\cite{Sun} Initially, the sample was cooled in 50 Oe field from 100 K ($T_{H}$) to the base temperature 2 K ($T_{base}$) at a constant cooling rate of 1 K/min. After reaching $T_{base}$, the sample was continuously heated back to $T_{H}$ at the same rate and the FC magnetization was recorded. The obtained $M(T)$  is the conventional FC magnetization curve and referred to as $M_{FCW}^{ref}$. Then the sample was cooled again at the same rate from 100  K to 2 K with intermediate stops of duration $t_{w}$=2 h each at $T_{stop}$=26, 18, and 10 K and the magnetization was also recorded during these halts. During each stop, the magnetic field was also turned off to let the magnetization decay. At each stop, after the lapse of $t_{w}$, the same field was reapplied and cooling was resumed. This cooling protocol results in a step-like $M(T)$ curve ($M_{FCC}^{stop}$). Finally, after reaching the base temperature 2 K, the sample was heated back continuously at the rate of 1 K/min in presence of 50 Oe field and the magnetization was recorded. Despite the continuous heating, the $M(T)$ curve obtained in this way exhibits a clear upturn at each $T_{stop}$, revealing the previous history of zero-field-relaxation at that $T_{stop}$ and resembles the previous step-like shape. This curve is referred to as the ``memory" curve ($M_{FCW}^{mem}$) which manifests clearly the memory effect in the temperature dependence of the FC magnetization for the present compound (Fig. 4). The FC memory effect $M_{FCW}^{mem}$ of this single-crystalline compound is quite similar to the numerically simulated FC memory curve for the interacting glassy system.\cite{malay}  It may be noted that the nature of the temperature dependence of FC magnetization below the freezing temperature for the present sample is very different from that one expects for a superparamagnetic system.   FC magnetization should increase with the decrease of temperature in superparamagnets.\cite{malay} \\
\begin{figure}
\includegraphics[width=0.5\textwidth]{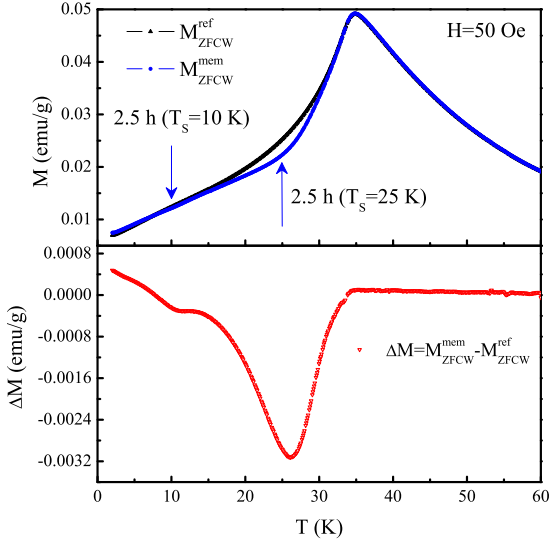}
\caption{(Color online) The memory effect in the temperature dependence of the zero-filed-cooled magnetization in 50 Oe field. In the upper panel the magnetization curves $M_{ZFCW}^{ref}$ and $M_{ZFCW}^{mem}$ are the reference and memory curves, respectively. The lower panel shows the difference curve $\Delta M$ ($=$$M_{ZFCW}^{mem}$$-$$M_{ZFCW}^{ref}$) exhibiting clearly the presence of memory dips at 25 K and 10 K. }
\label{Figure.5}
\end{figure}
The memory effect in the field-cooled process may also result from the independent relaxation of metastable phase clusters that could be formed in a phase separated or superparamagnetic system.\cite{Sasaki} Therefore, memory effect has been further investigated in the ZFC magnetization to discard such possibility. Here, the sample was cooled rapidly in zero field from 100 K to some selected stopping temperatures $T_{S}$=25 and 10 K, where the temperature was maintained for 2.5 h. After reaching the base temperature (2 K), the sample was heated back continuously at the rate of 1 K/min in 50 Oe field and magnetization was recorded during the heating. This $M(T)$ curve is designated as $M_{ZFCW}^{mem}$ in Fig. 5. The conventional ZFC magnetization ($M_{ZFCW}^{ref}$) at 50 Oe was also recorded. The $M_{ZFCW}^{mem}$ curve exhibits characteristic features at the point of stops which can be clearly seen in the difference curve $\Delta M$ ($=$$M_{ZFCW}^{mem}$$-$$M_{ZFCW}^{ref}$). $\Delta M$($T$) curve exhibits memory dips at 25 and 10 K. Unlike the independent relaxation of metastable phase clusters in a noninteracting superparamagnetic system, the presence of memory dips in the ZFC mode establishes that the glassy behavior in this compound is due to the cooperative spin-spin interactions.\cite{Sasaki} In a spin glass, the spin-spin correlation length grows during the stop even in  absence of field and causes memory dip in $M(T)$ curve during subsequent reheating. So, the observed memory effect in the ZFC magnetization further supports the spin glass phase in the present compound.\\
\\
\begin{figure}
\includegraphics[width=0.5\textwidth]{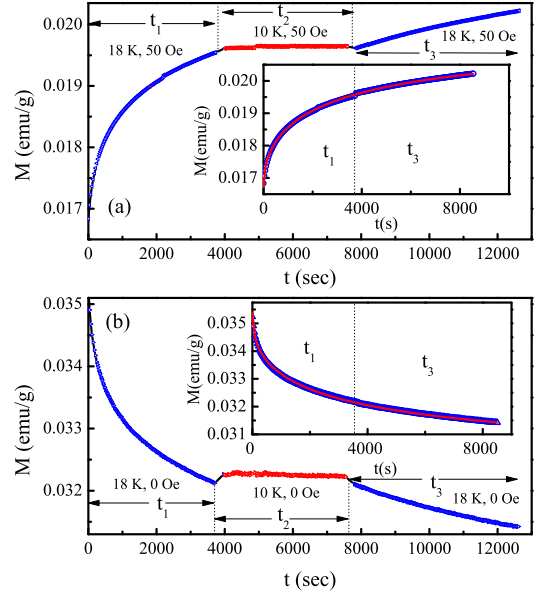}
\caption{(Color online) Magnetic relaxation at 18 K  for $H$=50 Oe with temporary cooling at 10 K taken in (a) the ZFC method and (b) the FC method. The insets show the same data vs the total time spent at 18 K. The relaxation curve during $t_{3}$ is the continuation of the curve during $t_{1}$ and the solid lines are due to the best fit to Eq. (1).}
\label{Figure.6}
\end{figure}
The memory effects in different glassy systems can be described well by two theoretical models viz., the droplet \cite{Fisher,Yoshino} and hierarchial models\cite{Evincent,Dotsenko,Lefloch,Vincent}. According to the droplet model, at a given temperature, only one equilibrium spin configuration (and its spin-reversed counterpart) exists whereas a multivalley  structure is formed on the free-energy landscape in the hierarchical model. The critical difference between these two scenarios is the presence or absence of original spin configuration  after a positive temperature cycling. In droplet model, the original spin configuration is restored upon temporary heating as well as cooling whereas according to the hierarchical model a temporary heating  rejuvenates the relaxation process and there is no memory effect. So to test these arguments as well as to study the memory effects in the time variation of magnetization, we have investigated the influence of temperature and field cycling on the behavior of magnetic relaxation in both  FC and ZFC methods following the protocol of Sun \emph {et al.}\cite{Sun} Figs. 6(a) and 6(b) display the magnetic relaxation behavior with temporary cooling under ZFC and FC conditions, respectively. In the ZFC process, the sample was cooled down to a temperature $T_{0}$=18 K from 100 K at a cooling rate of 1 K/min in  absence of field. After reaching $T_{0}$,  50 Oe field was applied and the magnetization was recorded as a function of time ($t$). After the lapse of a time period $t_{1}$=1 h, the sample was quenched in the constant field to a lower temperature $T_{0}$$-$$\triangle T$=10 K and magnetization was recorded for time period $t_{2}$=1 h. After the lapse of $t_{2}$, the sample was heated back to $T_{0}$ and magnetization was recorded for another time $t_{3}$=1 h. From Figs. 6(a) and 6(b), one can see that during the temporary cooling the relaxations become very weak and when the temperature returns to $T_{0}$, the magnetization comes back to the level it had reached just before the temporary cooling. The insets show that the relaxation during $t_{3}$ is the continuation of that during $t_{1}$ and the entire relaxation during $t_{1}$ and $t_{3}$ can be fitted to a single curve following Eq. (1). To measure the strength of the memory phenomena in the time dependence of magnetization, we have allowed the sample to undergo opposite relaxations by switching on and off the applied field in the FC and ZFC methods, respectively, during time interval $t_{2}$. It is noteworthy that despite such opposite relaxation during $t_{2}$, the magnetic relaxations during time interval $t_{3}$ are almost the same as that during $t_{1}$ as shown in the Figs. 7(a) and 7(b). Again, the relaxations during $t_{1}$ and $t_{3}$ can be described well by a single curve following Eq. (1) as shown in the insets of Figs. 7(a) and 7(b). The restoration of the original spin configuration even after a large change in magnetization ($\sim$88\% and $\sim$40\% for the ZFC and FC modes, respectively) due to field cycling suggests that the memory effect in this compound is indeed quite strong. Therefore, the above results clearly demonstrate the striking memory effects in LSCO single-crystalline compound.\\
\begin{figure}
\includegraphics[width=0.5\textwidth]{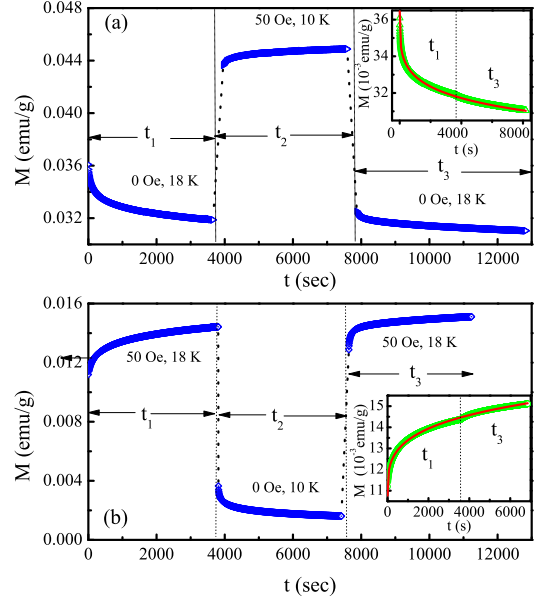}
\caption{(Color online)  Magnetic relaxation at 18 K with an opposite relaxation during temporary cooling at 10 K for (a) the ZFC method and (b) the FC method. The insets plot the same data vs the total time spent at 18 K. The relaxation curve during $t_{3}$ is the continuation of the curve during $t_{1}$ and the solid lines are the best fit to Eq. (1).}
\label{Figure.7}
\end{figure}
\begin{figure}
\includegraphics[width=0.5\textwidth]{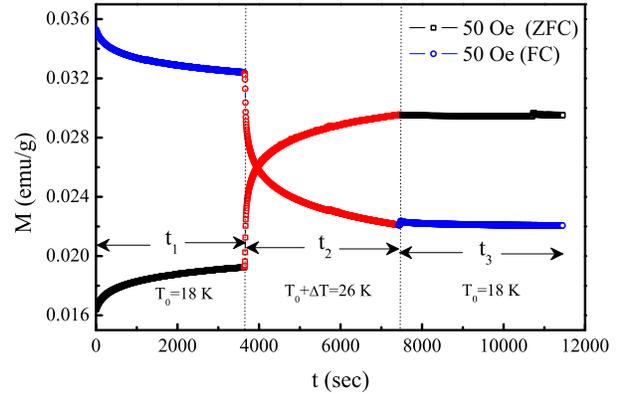}
\caption{(Color online) Magnetic relaxation at 18 K with an intermediate heating at 26 K taken in the ZFC and FC modes for $H$=50 Oe. Positive temperature cycling rejuvenates the relaxation at the higher temperature and no memory effects appear after temperature returns.}
\label{Figure.8}
\end{figure}
The effect of positive temperature cycling in the magnetic relaxation has also been investigated and is shown in Fig. 8. Both FC and ZFC methods have been used. In the ZFC method, the sample was cooled from 100 K to $T_{0}$=18 K in absence of field and then magnetization was recorded for time interval $t_{1}$=1 h after applying 50 Oe magnetic field. After the lapse of $t_{1}$, the sample was temporarily heated to $T_{0}$$+$$\triangle T$=26 K and the magnetization was recorded for time period $t_{2}$=1 h. Finally, the sample was cooled back to initial temperature $T_{0}$=18 K and magnetization was recorded for another time period $t_{3}$=1 h. Here, we see that unlike negative temperature change the magnetization at the beginning of $t_{3}$ does not come back to the level it had reached just before the temporary heating and the nature of magnetic relaxation during time period $t_{3}$ is quite different from that during $t_{1}$. Similar behavior has also been observed in the FC method. So, the temporary heating rejuvenates the magnetic relaxation and hence no memory effect in the positive temperature cycling. Therefore, the asymmetric response with respect to negative and positive temperature change favors the hierarchial model of the relaxation for the present compound.
\\

The above studies on magnetic relaxation and various memory effects support the spin glass phase in this single-crystalline compound. In manganites where the FM clusters are embedded in the antiferromagnetic matrix, the glassy dynamics can be explained by considering only the interaction among these FM clusters.  This inter-cluster interaction can also be tuned by composition and/or magnetic field due to the change in the size and concentration of the magnetic clusters which determines the frustration and the collectivity observed in the relaxation of the system.\cite{Rivadulla} But the spin-spin interaction/frustration in cobaltite that gives rise magnetic relaxation below $T_{f}$ is independent of temperature, magnetic field as well as doping concentration similar to that of a true spin glass system.  The features of spin glass phase have also been reflected in the memory effects observed in this compound. The asymmetric response of the magnetic relaxation due to positive and negative temperature change points toward the presence of hierarchical organizations of the metastable states in the glassy phase. In the hierarchical picture of spin glass, at a given temperature ($T_{0}$), numerous valleys (metastable states) are formed on the free-energy surface. When the system is quenched from $T_{0}$ to $T_{0}-\triangle T$ each free energy valley splits and forms a set of new smaller subvalleys. For large enough $\triangle T$, it is not possible to overcome the barriers separating the main valleys during the relaxation time $t_{2}$  and the relaxation occurs only within the newly born subvalleys of each set. When the sample is bring back to $T_{0}$, these new subvalleys and barriers merge back to the original free-energy landscape so that the relaxation during the temporary cooling does not contribute to the relaxation at $T_{0}$. This explains the observed memory effects during temporary cooling. In temporary heating period, the magnetic relaxation restarts in a fresh landscape created by the merging of the valleys formed at lower temperature $T_{0}$. When the sample is cooled back to $T_{0}$ after the end of heating period, the system is trapped in one of the redivided valleys in the region explored during temporary heating. The restoration of original spin configuration at $T_{0}$ is then seems improbable as there are many redivided valleys other than the original. It is worth mentioning that even a considerable field change (50 Oe) during temporary cooling does not affect the memory effect significantly, implying that this hierarchical configuration is robust in LSCO compound.\cite{Sun} Recently, some studies have been performed on the universality of spin configuration restoration in metallic and insulating Heisenberg spin glasses.\cite{hMamiya,Mamiya} Those results demonstrate that memory effects are not attributable to the preservation of a simply frozen state but to the spontaneous restoration of the original spin configuration. The present study  also supports this observation. However, unlike the droplet model that discusses the memory effects in others metallic and/or insulating Heisenberg spin glasses, the hierarchical model holds good for the present system. Further studies on similar systems should be performed to shed some more light on this issue.\\

\section{Conclusion}

We have presented a comprehensive study of magnetic relaxation and memory effects in La$_{0.9}$Sr$_{0.1}$CoO$_{3}$ single crystal. The magnetic relaxation can be described well by the stretched exponential function and shows that the system evolves through a number of intermediate states. The analysis of the magnetic relaxation rate at different temperatures and magnetic fields shows that the glassy behavior of this single-crystalline compound resembles that of a true spin glass phase and is quite different from that observed in manganites where only the inter-cluster interaction is the origin of the glassy behavior. The observed spin glass behavior in the present studies believed to be due to the random distribution of ferromagnetic and antiferromagnetic interaction in the system. Memory effects have been clearly observed in different temperature and field cycling experiments and show that the compound is capable of retaining the magnetization history even for a large change in the magnetization. The presence of memory dips in the temperature dependence of the ZFC magnetization rules out any possibility of superparamagnetic like behavior originating from independent relaxation of the metastable phase clusters. The effects of positive and negative temperature changes on the reversion of original spin configuration suggest that the memory phenomena in this compound follow the hierarchical model of spin glass.
        \\
\section{Acknowledgement}

The authors would like to thank A. Pal for technical help during measurements.\\

\end{document}